\newcommand{\mytitle}{%
  \textsc{\textmd{CHESS}}: Cloud, High-Performance Computing, and Edge for Science and Security}

\documentclass[11pt]{article}

\usepackage{cfg-doc}

\newcommand{\mybibstyle}{abbrv}

\usepackage{cfg-me}

\titlespacing{\section}{0pt}{1.3ex}{1ex}

\titlespacing{\subsection}{0pt}{1ex}{0ex}

\titlespacing{\subsubsection}{0pt}{0.5ex}{0ex}

\titlespacing{\paragraph}{\parindent}{0.1ex}{1ex}
\newcommand{\mychaptertitle}[1]{{\fontfamily{cmss}\fontseries{mc}\selectfont\Large\noindent\color{black}{\MakeUppercase{#1}}\\[-0.5em]\noindent\makebox[\linewidth]{\rule{\textwidth}{0.4pt}}}}

\counterwithin{figure}{section}
\counterwithin{table}{section}

\colorlet{shadecolor}{pastelblue}%lavendermist,pastelblue
\newlist{mytasklist}{enumerate}{1} % 'enumerate*' is inlined
\setlist[mytasklist]{label=(\alph*),nosep}

\begin{document}

%% Outline
%% \renewcommand*\contentsname{Overview (for draft; 16 pg max)} % \phantom{}
%% \tableofcontents
%% \clearpage

%============================================================================
% 
%============================================================================

%\pagestyle{empty}

\title{Final Report for \mytitle}

\author{%
  Nathan Tallent
  \and
  Jan Strube
  \and
  Luanzheng Guo
  \and
  Hyungro Lee
  \and
  Jesun Firoz
  \and
  Sayan Ghosh
  \and
  Bo Fang
  \and
  Oceane Bel
  \and
  Steven Spurgeon
  \and
  Sarah Akers
  \and
  Christina Doty%\\
  \and
  Erol Cromwell
}

\date{%
  \vspace{20em}%
  Technical Report PNNL-36859%
  \endgraf\bigskip
  Pacific Northwest National Laboratory
  \endgraf\vspace{5em}%
  October 14, 2024%
}

% \maketitle sets page style to plain so move \thispagestyle{empty}
\clearpage
\maketitle
\thispagestyle{empty}

\newpage
\setcounter{page}{1}

%============================================================================
%============================================================================

\section{Executive Summary: \\ Automation for Autonomous Science and Continuum Platforms}
\label{sec:abstract}

Impactful science increasingly requires collaborations combining instruments and datasets at multiple facilities.
Consider the formation of digital twins of complex physical systems; or the automation of experimental facilities.
A core challenge is automation of the underlying theory-experiment cycle that consists of (a) guidance and explanation from theory and (b) measurement and validation from experiments~\cite{nas:workflows:2022,Ferreira:2024:Computer-workflow-frontiers,FerreiraDaSilva:2024:WorkflowsSummit}.

\begin{wrapfigure}[20]{r}{0.5\linewidth}%
%\begin{figure}[b]
  \centering
  \vspace{-1.5ex}
  \includegraphics[width=1.0\linewidth]{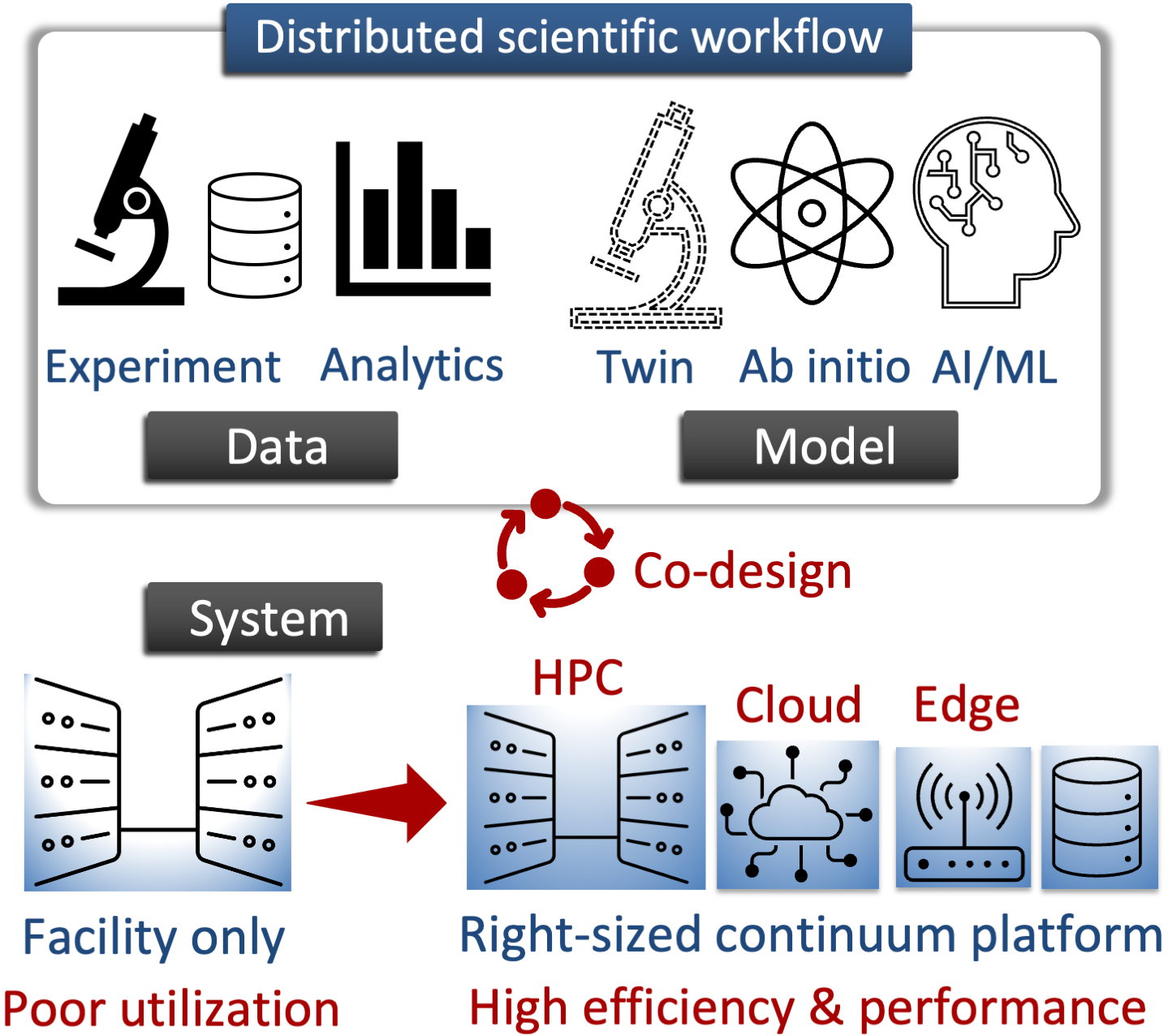}
  \caption{Advancing collaborative computational science by automating distributed scientific systems on continuum platforms.}
  \label{fig:chess-overview-abstract}
%\end{figure}
\end{wrapfigure}%

Automating the theory-experiment cycle requires effective distributed workflows that utilize a computing continuum spanning lab instruments, edge sensors, computing resources at multiple facilities, data sets distributed across multiple information sources, and potentially cloud.
Unfortunately, the obvious methods for constructing continuum platforms, orchestrating workflow tasks, and curating datasets over time fail to achieve scientific requirements for performance, energy, security, and reliability.
Furthermore, achieving the best use of continuum resources depends upon the efficient composition and execution of workflow tasks, i.e., combinations of numerical solvers, data analytics, and machine learning.

Pacific Northwest National Laboratory's LDRD \emph{Cloud, High-Performance Computing (HPC), and Edge for Science and Security} (CHESS) has developed a set of interrelated capabilities for enabling distributed scientific workflows and curating datasets. 
CHESS has 
\begin{myitemize}

\item Blazed a transition path for scientists to explore continuum computing with cloud using portable templates and goal-directed executions that can optimize for performance and dollar cost.

\item Created AI-aware services for workflows and data management, including
multi-modal LLM pipelines;
error-bounded multi-modal dimensionality reduction;
guided and goal-directed model search (performance, accuracy);
microstructure-aware image segmentation;
microstructure-aware data compression;
high-performance training for graph neural networks;
and novel federated LLM training.

\item Designed workflow measurement, modeling, prediction, and scheduling for co-design of continuum computational science composed of data-driven models, physical simulations, and data curation.

\item Demonstrated impact:
  enabling portable, goal-directed continuum computing;
  improving distributed workflow response time (1.28× -- 87× speedup);
  creating high-performance microstructure-aware labeling (+17\% absolute accuracy; -14\% reduction of false positives) and compression (10-12× speedup);
  creating AI-based distributed services, training for science.

\end{myitemize}

%============================================================================
%============================================================================

\newpage

\section{Introduction}
\label{sec:introduction}

Pacific Northwest National Laboratory's ``Cloud, High-Performance Computing (HPC), and Edge for Science and Security'' (CHESS)~\cite{chess-www} is a laboratory directed research and development effort to build capabilities to support integrated computing plans across three very different computing environments:
\begin{itemize}
\item \emph{Cloud computing} --- data is processed at a different location than the originating source.
\item \emph{High-performance computing} --- complex data processing calculations can be completed at a high rate.
\item \emph{Edge computing} --- data is processed as close to the originating source as possible.  
\end{itemize}
The vision is to create tooling and automation that allows researchers to seamlessly bridge these three environments to effectively create a virtual and distributed supercomputer that retains all the benefits of the individual environments, but then enhances these by allowing compute workloads to automatically and efficiently move between these environments to leverage additional computing, or other services not readily available in the current environment.

%\begin{wrapfigure}[20]{r}{0.5\linewidth}%
\begin{figure}[h]
  \centering
  \includegraphics[width=0.8\linewidth]{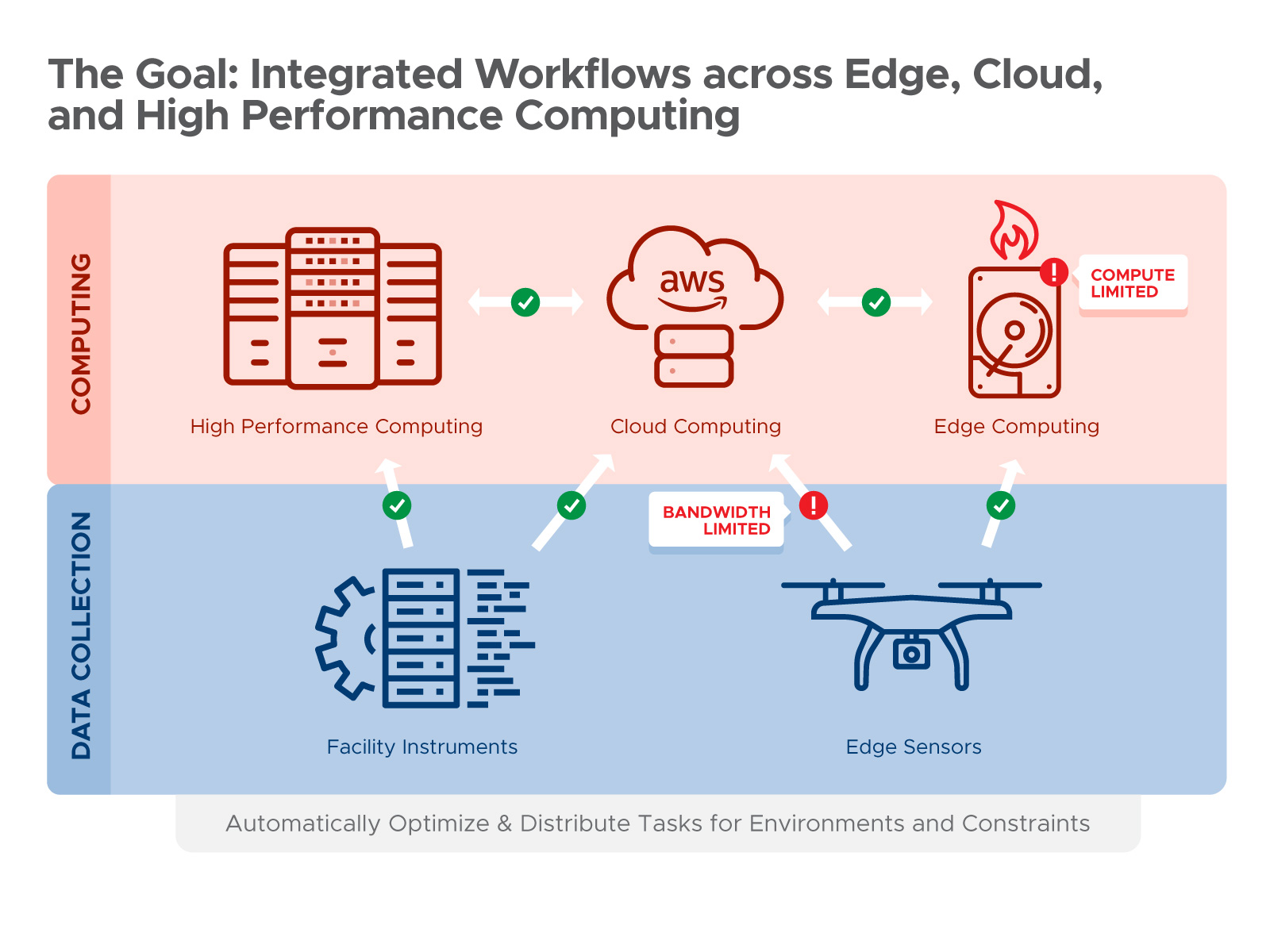}
  \caption{PNNL is building workflows that span the spectrum of compute environments and with strategies for moving data and AI workloads between them. CHESS tests these strategies against use-cases which require moving models between such environments.}
  \label{fig:chess-overview-web}
\end{figure}
%\end{wrapfigure}%

The project consists of two major thrusts, Security and Science. %, and Foundational Cloud Capabilities.
This report focuses on the latter.

%===========================================================
%===========================================================

\subsection{CHESS' Science Thrust}
%\myparagraph{CHESS' Science Thrust}

New materials have the potential for improving solar generation, creating new batteries, developing new health care treatments, and enabling new techniques in computing. The problem is that new materials with just the right properties are extremely hard to find.

\begin{wrapfigure}[18]{r}{0.5\linewidth}%
%\begin{figure}[b]
  \centering
  \vspace{-1.5ex}
  \includegraphics[width=1.0\linewidth]{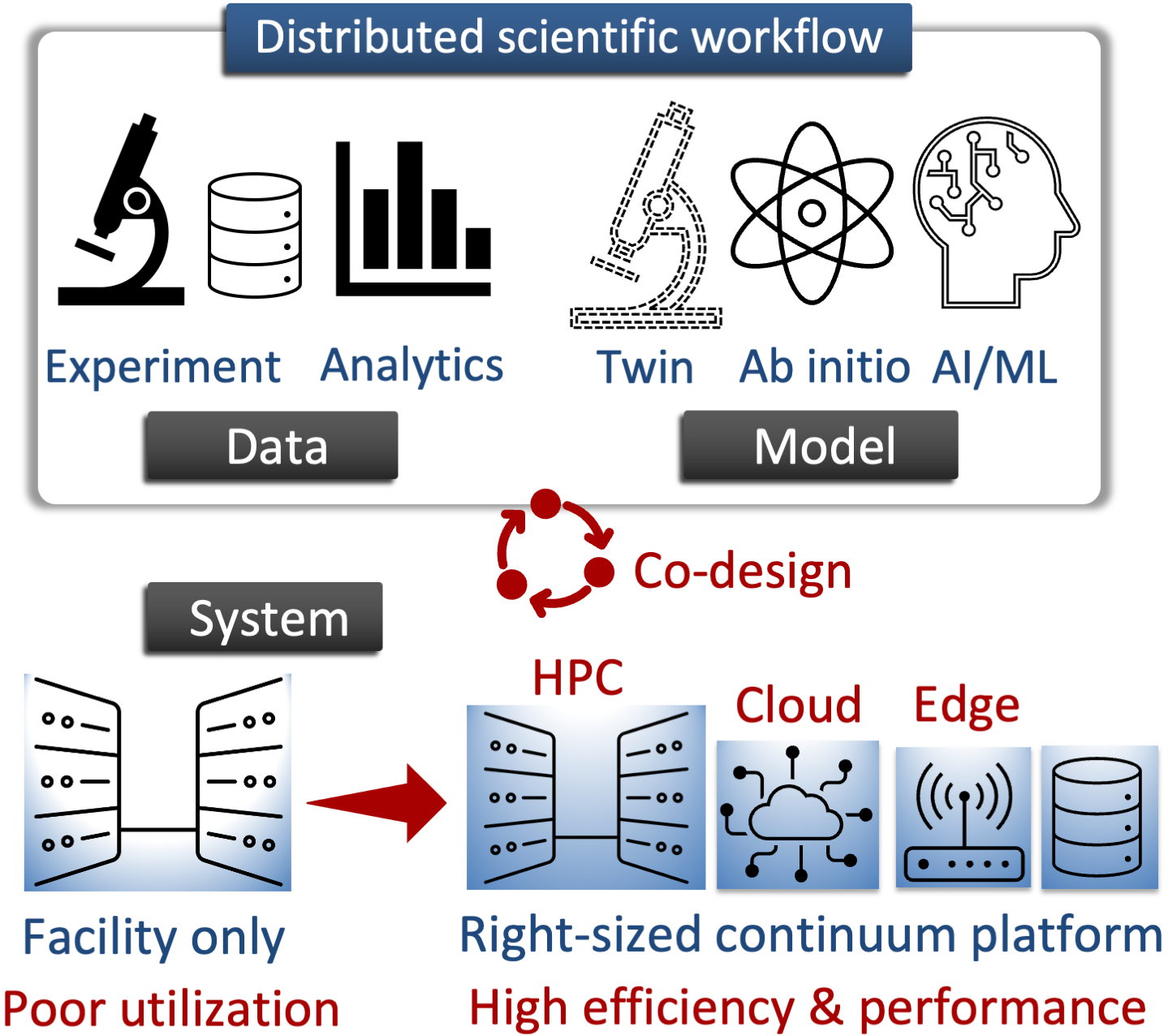}
  \caption{CHESS's research targeted co-design distributed scientific systems for AI-enabled computational science.}
  \label{fig:chess-overview}
%\end{figure}
\end{wrapfigure}%

The key to accelerating scientific discovery is automation of the complex theory-experiment cycle that consists of (a) guidance and explanation from theory and (b) measurement and validation from experiments~\cite{nas:workflows:2022,Ferreira:2024:Computer-workflow-frontiers,FerreiraDaSilva:2024:WorkflowsSummit}.

As straightforward automation methods simply cannot deliver responses within required parameters (time, quality, cost, resilience, etc), new computational and control methods are needed for controlling and coordinating large computational models, hypothesis generation, instrument control, experimental interpretation, feedback, data movement, and datasets.

% From a computing automation perspective, the implication
More broadly, collaborative science involving instruments and datasets at multiple facilities is increasingly important.
The implication is that these collaborations require executing distributed workflows in a computing continuum that spans lab instruments, edge sensors, computing resources at multiple facilities, data sets distributed across multiple information sources, and potentially cloud.
Unfortunately, the obvious methods for constructing a continuum platform and of orchestrating workflow tasks usually fail to achieve scientific requirements for performance, energy, security, and reliability.
Furthermore, achieving the best use of continuum resources and coordination depends upon efficient composition and execution of its constituent tasks, i.e., combinations of numerical solvers, data analytics, and machine learning.

\begin{wrapfigure}[16]{r}{0.5\linewidth}%
%\begin{figure}[b]
  \centering
  \vspace{-2ex}
  \includegraphics[width=1.0\linewidth]{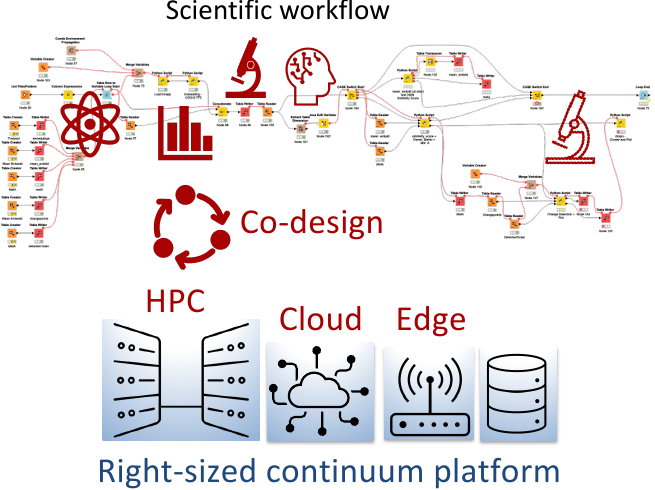}
  \caption{CHESS's research challenge.}
  \label{fig:chess-challenge}
%\end{figure}
\end{wrapfigure}%

CHESS's science thrust is developing the computing and coordination techniques for automating theory experiment cycles.
New methods are needed for representing and executing the computational workflows that are formed between instruments and computational devices ranging from near-instrument devices, high performance computing systems, and specialized demand-driven cloud resources.
\Cref{fig:chess-overview} shows CHESS's goal of co-design between Data-driven methods, Models, and Continuum computing systems.
CHESS's research is exploring techniques for productive and portable representations, scheduling and resource assignment, diagnosing and removing performance bottlenecks, AI/ML methods for domain semantics, and customized data compression to adjust data flow.

\Cref{fig:chess-challenge} shows the challenge.
The composition of data-driven methods, models, and geo-distributed datasets creates distributed scientific workflows, represented using the complex graph at the top.
For example, the co-design process must determine the best composition of workflow subtasks and their placement over a continuum platform.

%===========================================================
%===========================================================

\subsection{Results}

\begin{figure}[tb]
  \centering
  \includegraphics[width=1.0\linewidth]{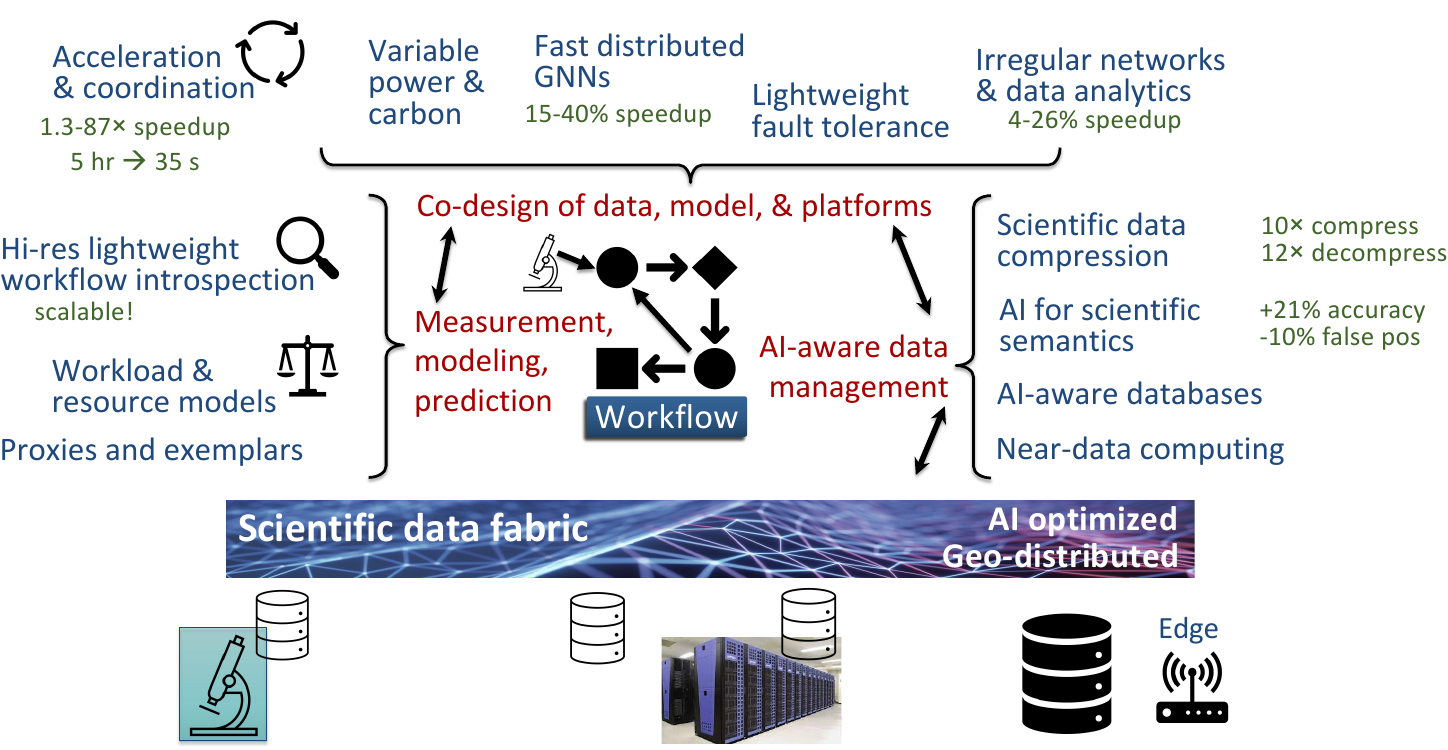}
  \caption{Overview of CHESS's science results, centered around (a) measurement, modeling, prediction that enables (b) co-design of continuum science consisting of data-driven methods, models and simulations, and continuum platforms. Additionally we explore (c) data management techniques with a view towards AI-aware data-driven science.}
  \label{fig:chess-results}
\end{figure}

CHESS science thrust has developed a set of interrelated capabilities for co-design between distributed science and continuum platforms.
An overview of the capabilities and associated results is shown in \cref{fig:chess-results}.
The thrust's focus has been primarily on (a) measurement, modeling, prediction that enables (b) co-design of continuum science consisting of data-driven methods, models and simulations, and continuum platforms.
Additionally we explore (c) data management techniques with a view towards AI-aware data-driven science.

The former consists of high-resolution but lightweight workflow introspection and models for workloads and resources.
The latter consists of methods for improving coordination via scheduling tasks and assigning tasks to resources.
Note that this includes improving data layouts and schedules for data movement.
The the most important methods are described in \cref{sec:artifacts:details}.

The results of these methods are as follows.

\begin{figure}[t]
  \centering
  \includegraphics[width=0.5\linewidth]{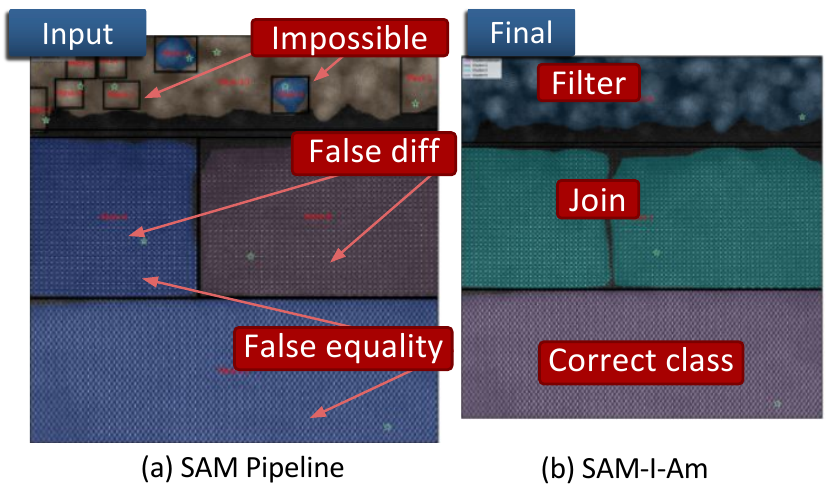}
  \caption{CHESS developed SAM-I-Am, new \emph{microstructure segementation} for interpreting images from STEM microscopy.}
  \label{fig:samiam}
\end{figure}

\myparagraph{Co-design and Capability transfer}.
\begin{myitemize}
\item Blazed a transition path for scientists to explore continuum computing with cloud using portable templates and goal-directed executions that can optimize for performance and dollar cost.
  Any plausible transition path cannot require that scientists rewrite their whole environmental assumptions. % when we know that compatible AWS solutions exits.
  We helped to define PNNL's ``Cloud Operations Council'' that includes researchersp, C3, Security, and CloudOps to identity and prioritize the enhancements that were needed to enable our core science mission.
  
\item Created AI-aware services for workflows and data management, including
multi-modal LLM pipelines;
error-bounded multi-modal dimensionality reduction;
guided and goal-directed model search (performance, accuracy);
microstructure-aware image segmentation;
microstructure-aware data compression;
high-performance training for graph neural networks;
and novel federated LLM training.

\item Designed workflow measurement, modeling, prediction, and scheduling for co-design of continuum computational science composed of data-driven models, physical simulations, and data curation.

\item Demonstrated impact:
  enabling portable, goal-directed continuum computing;
  improving distributed workflow response time (1.28× -- 87× speedup);
  creating high-performance microstructure-aware compression and labeling (+17\% absolute accuracy; -14\% reduction of false positives) and shown in \cref{fig:samiam};
   and compression (10-12× speedup);
  creating AI-based distributed services, training for science.
    
\end{myitemize}

%\begin{wrapfigure}[20]{r}{0.5\linewidth}%
%\end{wrapfigure}%

\myparagraph{Measurement, modeling, prediction}.
We have introduced efficient fine-grained performance bottleneck detection~\cite{Lee+:2023-SC:data-flow-lifecycles,Tang+:2024:CLUSTER-dayu} that informs novel linear-time scheduling has demonstrated 1.28×, 87×, 1.4×, and 5× speedups on common workflows~\cite{Guo+:20xx:fast-flow,Hou+:2024:ISNCC-near-data}.
A compression-enabled roofline model is proposed by delicately enabling compression and approximation into computation~\cite{Soumya+:20xx:compression-roofline}. This new model aims to balance and transition between computational and memory demands, mitigate and transform performance bottlenecks, and dynamically adjust to the specific strengths and limitations of the underlying hardware.
We have developed models for understanding and predicting cross-application I/O interference in distributed storage systems~\cite{Egersdoerfer-Fang+:20xx:cross-app-io-interference}.
Novel data-movement has achieved 10× on the important BelleII MC workflow \cite{Lee+:2023-SC:data-flow-lifecycles} (\cf \cite{Friese:2020:BigData-tazer-io-composition,Suetterlein:2019:BigData-tazer}).
Understanding workflow-wide semantic data flow storm tracking pinpointed unexpected dataset overheads of 3× on an E3SM-companion storm tracker~\cite{Tang+:2024:CLUSTER-dayu}.

\myparagraph{Co-design and data driven analytics}.
For distributed training, we introduced distributed training into XCT (X-ray computed tomography) X-Ray image interpretation, accelerating the training time by orders of magnitude (5 hr to 35 s).
Further, MassiveGNN introduces performant and productive training for massively connected (distributed) GNNs, improving the state-of-the-art Amazon DistDGL by 15--40\% \cite{Sarkar+:2024:CLUSTER-massive-gnn}.

We developed new techniques for accelerating data flow in data-intensive HPC and graph analytics.
We developed network-agnostic locality-aware process assignment heuristics for distributed-memory graph workloads %by exploiting the structure of input graphs
and demonstrated 30--40\% improvements in communication times in simulation and practice.
% For four communication intensive distributed-memory graph workloads --- Breadth First Search (BFS), Louvain Clustering, Triangle Counting and Single Source Shortest Path (SSSP), we demonstrate up to about 30--40\% improvements in the overall MPI communication times through proposed process remapping methodologies via packet-level simulations using Structural Simulation Toolkit (SST) primarily on Dragonfly based network topology and validate the strategies empirically on HPE Slingshot network of the NERSC Perlmutter supercomputer.
We demonstrated that unstructured topologies, compared to popular Dragonfly and Fat Tree, not only benefit performance of communication-intensive HPC applications and graph analytics, but are now practical \cite{Newaz+:2024:IPDPS-jellyfish-graph-analytics}.
% Explain, using detailed simulations, that iso-cost Jellyfish topologies have more diversity of paths, resulting in sustainable performance improvements. Show how to configure Jellyfish to achieve cost parity vs. Dragonfly and Fat Tree.

\myparagraph{Co-design and domain science}.
Co-design with application scientists resulted improved techniques for scientific image interpretation.
A new material interpretation model introduces vastly improved material identification for STEM (scanning transmission electron microscopy) images (zero-shot accuracy increase of absolute 21\%, 13\%, 5\% and a false positive reduction of absolute 10\%, 18\%, 4\%)~\cite{Abebe+:20xx:SamIAm-semantic-boosting,Abebe+:2025:SamIAm-semantic-boosting}.
We created a SAM-I-Am-assisted labeling assistant within LabelStudio to to rapidly create precise labels for image segments of identical materials.
Before using this tool we could only obtain 3 labeled images in several hours; with it, we obtained 50 labeled images in under 2 hours.
We also release an online demonstration~\cite{samiam-demo}.

The same co-design has produced a new scientific image compressor for managing data flow and data sets.
Unlike other compressors, the ViSemZ image compressor both preserves material texture semantics and is highly performant (10-12× speedup)~\cite{Zhang+:20xx:visemz-semantic-compression}. 
Further, a large language model-based compression technique has been studied on multi-modal/-domain/-dimensional scientific data~\cite{Max+:20xx:llm-lossless-compression}.

\myparagraph{AI aware dataset management.}
With colleages at University of Washington, we have developed a dimensionality reduction theory motivated by the challenges of multi-modal data and similarity search~\cite{Gong+:20xx:opdir-order-preserving-dimreduce}.
We believe the theoretical insights will have important applications for of scientific data meshes such new scaling limits and search capabilities.

%============================================================================
%============================================================================

\section{Potential benefits of cloud for laboratory scientists}
\label{sec:cloud-potential}

Cloud now dominates the computing landscape and drives much of the innovation in hardware, system virtualization, and AI frameworks~\cite{Reed:2023:reinventing-hpc} \cite{nas:beyond-exascale:2023}.
Different U.S. government-related organizations have explored the potential of cloud and reached different conclusions~\cite{Mehrotra+:2016:hpc-in-cloud,Guidi:2020:hpc-cloud,ornl:2023:hpc-cloud}.
% NASA~\cite{Mehrotra+:2016:hpc-in-cloud}.
% NERSC~\cite{Guidi:2020:hpc-cloud}.
% ORNL~\cite{ornl:2023:hpc-cloud}.
CHESS' vision is the intelligent selection of the right combination of facility and cloud resources.
Since cloud is unfamiliar to many laboratory scientists this section discusses some of the reasons for considering the use of cloud.

First, cloud offers some resources that are simply not available in a high-end HPC facility.
Cloud providers now design their own processors and accelerators that are not available for purchase.
% Very easy to find: ARM, RISC-V
%
Additionally, cloud provides an extremely large range of platform diversity, enabling a vast array of selectable combinations of compute power, memory, networking,short and long-term storage.
Further, public and private cloud resources may offer the first access to emerging computing technology, ranging from data-flow architectures to experimental quantumn machines.
Finally, cloud routinely is the gateway for the latest AI/ML frameworks and services; as well as the most advanced foundation models and generative AI services.
% DataBricks, AWS Athena (hard to beat)

Second, cloud offers rapid on-demand scaling, as is well known.
When time is critical, it can (in most cases) become negligable at the expense of increased cost.
Thus, the batch wait times that are normal at at a facility can be controlled.

Third, cloud provides the potential to rapidly construct and understand the implications of proposed deployment.
Once the design is fixed, the result can be redeployed at the target target.
One common example is matching a collaborator's environment to validate a prototype and ensure compatibility on the target.
In some cases, a cloud vendors service difference may provide a way to quickly assemble and test an idea so that time can be spent on the new unexplored question rather than the known sub-components.

% Our position: Hybrid cloud + on prem. For example, HPC and gpu on prem; edge procesing on prem.

Two potential downsides of cloud are vendor lock and cost.
Both can be significant concerns, but should be understood in context.
Vendor lock is generally undesirable, but it may be worth the price if the alternative simply is not available or would require significant implementation efforts.
For costs, cloud costs vary substantially based on the utilized services and resources.
In general, one should carefully consider costs and bias designs toward less costly services.
At the same time, all costs must be considered.
If the alternative to cloud would require additional hardware, expertise/personnel, or implementation time, then cloud costs may be justified.

%============================================================================
%============================================================================

\section{Usage Scenarios and Tools}
\label{sec:artifacts}

This section describes different usage scenarios for a common workflow pattern, sketches templates for different execution objectives, and describes the artifacts and tools that CHESS has produced.

%===========================================================
%===========================================================

\subsection{Usage Scenarios and Templates}
\label{sec:artifacts:scenarios}

\newcommand{\mytask}[1]{\textsf{#1}}

\begin{wrapfigure}[12]{r}{0.3\linewidth}%
%\begin{figure}[b]
  \centering
  \vspace{-2ex}
  \includegraphics[width=1.0\linewidth]{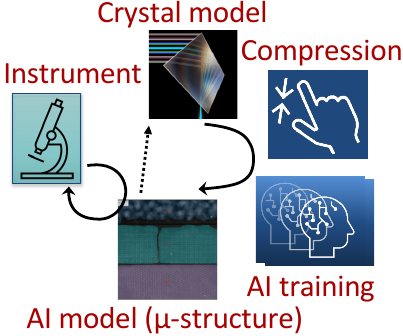}
  \caption{Overview of CHESS's workflow.}
  \label{fig:chess-workflow}
%\end{figure}
\end{wrapfigure}%

CHESS science thrust focused on instantiating a novel prototype scanning transmission electron microscopy (STEM) workflow that takes images from a STEM microscope, interprets them by precisely segmenting each distinct region using a co-designed technique called microstructure segmentation, and then (potentially) refine microscope settings based on the interpretation or guide the planning of subsequent material samples.
An overview of the workflow tasks is shown in \cref{fig:chess-workflow}.
The figure shows 5 primary tasks, each of which contains many sub-components.
The tasks form an \emph{inner} and \emph{outer} loop.
The inner loop is the core experiment-interpretation cycle, i.e., STEM image to interpretation to STEM adjustment.
Concurrenly, the ``outer loop'' executes at a slower frequency that generates synthetic image data using a multi-slice crystal model and updates the microstructure segmentation with distributed retraining.
The distributed retraining leverages dynamic data models to reducing expensive communication.
Finally, microstructure-aware compression --- a result of co-design --- is applied to manage the volumes and transfer rates of both the experimental and synthetic datasets.
Overall, this concurrent inner-loop and outer-loop pattern is encountered very frequently when automating the experiment-theory cycle.

Based on this general workflow pattern, we can consider different usage scenarios:
\begin{itemize}
\item optimize response time and cloud cost;
\item optimize cloud cost; and
\item optimize response time, energy, and cloud cost.
\end{itemize}

The result is three different continuum platforms, where each one is designed and instantiated based on the objective.
The experiment is designed to demonstrated the following:

\begin{myitemize}

\item The ability to vary optimization targets to achieve different goals.
  We exploit the uniqueness of each scenario, while grounding each scenario in typical constants such as a facility instrument, an ``inner decision loop'' for experimental interpretation, and an outer decision loop for inter-experiment feedback.
  
\item Co-design of each workflow and platform to achieve intelligent and cost-effective resource allocations based on understanding tasks and data flow.
  
\item Performant and portable orchestration that is compatible with open science % with NextFlow
\end{myitemize}

\begin{table}[h]% !htb
  \centering%
  {%\scriptsize
  %\resizebox{\linewidth}{!}{
  \setlength\tabcolsep{3.0pt}% default 6.0pt
\begin{tabular}{|p{0.15\linewidth}|p{0.85\linewidth}|}
  \hline%\toprule
  Task & Purpose \\
  \hline\hline%\midrule
  \mytask{Instrument} &
  STEM microscope experiments \\
  \hline
  \mytask{AI model} &
  Experimental interpretation with SAM-I-Am \\
  \hline
  \mytask{Crystal model} &
  Generate training data using first-principles method (PRISM, multi-slice modeling) \\
  \hline
  \mytask{AI Training} &
  Update AI model using (MassiveGNN) \\
  \hline
  \mytask{Compression} &
  Co-designed compression (ViSemZ) to preserve material textures for AI model \\
  \hline%\bottomrule
\end{tabular}
  \caption{Workflow tasks.}
  \label{tbl:workflow-tasks}
  }%
\end{table}

The specific tasks that compres the workflow in \cref{fig:chess-workflow} are described in \cref{tbl:workflow-tasks}.

\begin{itemize}
\item \mytask{Instrument}: STEM images, each with 2--4 materials from among compounds including La:SrTiO$_3$, SrTiO$_3$, Ge, Nb:SrTiO$_3$, WO$_3$, LaFeO$_3$, LaMnO$_3$, La$_{0.8}$Sr$_{0.2}$FeO$_3$, and Pt / C captured in annular dark-field (ADF) and high-angle annular dark-field (HAADF) modes.

\item \mytask{AI model}: Using SAM-I-Am, applies high-quality microstructure segmentation to STEM images in order to accurately distinguish between materials and (prospectively) label with the expected material formula, orientation, and STEM setting.
  
\item \mytask{Crystal model}: Given a description of the material's crystalline structure and microscope parameters, PRISMScope generates a range of (2D) image projections of a material (from a 4D original).
  The results enables association between a material, its crystalline structure, and STEM settings.
  
\item \mytask{AI Training}: MassiveGNN introduces performant and productive training for massively connected (distributed) GNNs within the state-of-the-art Amazon DistDGL distributed GNN framework.

\item \mytask{Compression}: Because the \mytask{Crystal model} generates large datasets, compression is valuable for minimizing data transfer costs and managing dataset volumes over time.
  ViSemZ is a high-performance image compressor (10-12× speedup) that preserves \emph{vi}sual \emph{sem}antics.

\end{itemize}

The sections below describe each scenario.
In each scenario two tasks are constant: \mytask{Instrument} and \mytask{AI model}.
The \mytask{Instrument} is located at a facility and cannot be moved.
The \mytask{AI model} executes on a GPU co-located with the \mytask{Instrument} to minimize response time and maximize control possibilites.
The other three tasks vary their configurations depending upon the optimization target.
  
We implement each scenario with the NextFlow~\cite{nextflow-url} workflow framework, described in \cref{sec:nextflow-intro}.
% We use NextFlow  as a high-quality exemplar of a framework.
Although there are other high-quality workflow frameworks, very few have the critical propreties that we judge to be important for open-science.
In particular we considered only frameworks that: enable productivity, performance, and portability; support multiple cloud vendors; retain local ownership and control; are customizable; are designed for open science; are open source; and are mature.
This is discussed further in \cref{sec:nextflow-intro}.

%=======================================
%=======================================

\subsubsection{Optimize for response time and cloud cost}
\label{sec:artifacts:scenarios:1}

To optimize for response time and cloud cost, we use the configuration in \cref{tbl:workflow-config-time-cost}.
This configuration demonstrates two important points.

First, it demonstrates the ability to select a different virtualization and availability model on cloud.
Cloud offers effective inexpensive spot instances to generate training data.
Because of alignment of need with cloud spot, there are minimal negative impacts.

Second, it demonstrates the ability to select exactly the right platform for parallel training \emph{and} optimize for price.
Most HPC machines have limited memory, meaning significant resources are wasted by scaling nodes to have sufficient memory.
The right resources leads to high utilization and efficiency.

We show how to quickly optimize for price using the SkyPilot (\cref{sec:skypilot-intro}) launching framework within NextFlow.
SkyPilot is an inter-cloud broker that can select the best resources both among multiple cloud providers and within a single clouddeach provider.
(The availability of computational resources and thier prices not only vary by cloud provider, but also within the regions of a single provider.)
Based on a task/workflow and resource specification, it selects the the resources that best meet the objective.

\begin{table}[h]% !htb
 \centering%
  {%\scriptsize
  %\resizebox{\linewidth}{!}{
  \setlength\tabcolsep{3.0pt}% default 6.0pt
\begin{tabular}{|p{0.15\linewidth}|p{0.85\linewidth}|}
  \hline%\toprule
  Task & Configuration \\
  \hline\hline%\midrule
  \mytask{Instrument} &
  Facility
  \\
  \hline
  \mytask{AI model} &
  Pair with microscope at facility to exploit near-data computing
  \\
  \hline
  \mytask{Crystal model} &
  Cloud/spot (Fargate) to exploit scaling of many independent jobs, price, and relaxed needs for full resilience
  \\
  \hline
  \mytask{AI Training} &
  Use cloud to obtain HPC resources unavailable at
  facility, specifically large-memory nodes with efficient
  interconnect. Transfer only updated models to facility. Explore
  potential of cloud/spot: vary availability (spot) and price (e.g.,
  AWS region), possibly with skypilot
  \\
  \hline
  \mytask{Compression} &
  Execute only for data to be downloaded from cloud
  \\
  \hline%\bottomrule
\end{tabular}
  \caption{Configuration to optimize response time and cloud cost.}
  \label{tbl:workflow-config-time-cost}
  }%
\end{table}

%=======================================
%=======================================

\subsubsection{Optimize for cloud cost}
\label{sec:artifacts:scenarios:2}

This experiment demonstrates the ability to minimize use of cloud, assuming HPC should be local.
The configuration is shown in \cref{tbl:workflow-config-cost}

\begin{table}[h]% !htb
 \centering%
  {%\scriptsize
  %\resizebox{\linewidth}{!}{
  \setlength\tabcolsep{3.0pt}% default 6.0pt
\begin{tabular}{|p{0.15\linewidth}|p{0.85\linewidth}|}
  \hline%\toprule
  Task & Configuration \\
  \hline\hline%\midrule
  \mytask{Instrument} &
  Same as \cref{sec:artifacts:scenarios:1} (\cref{tbl:workflow-config-time-cost}).
  \\
  \hline
  \mytask{AI model} &
  Same as \cref{sec:artifacts:scenarios:1} (\cref{tbl:workflow-config-time-cost}).
  \\
  \hline
  \mytask{Crystal model} &
  Combination of facility and cloud bursting: use cloud only when necessary to minimize cloud usage/cost. Cloud as in \cref{sec:artifacts:scenarios:1} (\cref{tbl:workflow-config-time-cost}).
  For data generated at facility, upload data to cloud.
  \\
  \hline
  \mytask{AI Training} &
  Use facility HPC to minimize cloud.
  \\
  \hline
  \mytask{Compression} &
  Same as \cref{sec:artifacts:scenarios:1} (\cref{tbl:workflow-config-time-cost}), but applies slightly differently.
  \\
  \hline%\bottomrule
\end{tabular}
  \caption{Configuration to optimize cloud cost.}
  \label{tbl:workflow-config-cost}
  }%
\end{table}

%=======================================
%=======================================

\subsubsection{Optimize for time, energy efficiency, and cloud cost}
\label{sec:artifacts:scenarios:3}

This experiment demonstrates the ability to additionally optimize for energy efficiency.
As all platforms are right-sized, wasted utilization and wasted energy (from poorly utilized compute and network) is minimized.
The configuration is shown in \cref{tbl:workflow-config-time-energy-cost}.

\begin{table}[h]% !htb
 \centering%
  {%\scriptsize
  %\resizebox{\linewidth}{!}{
  \setlength\tabcolsep{3.0pt}% default 6.0pt
\begin{tabular}{|p{0.15\linewidth}|p{0.85\linewidth}|}
  \hline%\toprule
  Task & Configuration \\
  \hline\hline%\midrule
  \mytask{Instrument} &
  Same as \cref{sec:artifacts:scenarios:1} (\cref{tbl:workflow-config-time-cost}).
  \\
  \hline
  \mytask{AI model} &
  Same as \cref{sec:artifacts:scenarios:1} (\cref{tbl:workflow-config-time-cost}) but consider time (resource availability)
  \\
  \hline
  \mytask{Crystal model} &
  Same as \cref{sec:artifacts:scenarios:2} (\cref{tbl:workflow-config-cost}).
  \\
  \hline
  \mytask{AI Training} &
  Similar to \cref{sec:artifacts:scenarios:1} (\cref{tbl:workflow-config-time-cost}) but smallest most efficient nodes yielding highest utilization. % contrast with \cref{sec:artifacts:scenarios:2}.
  \\
  \hline
  \mytask{Compression} &
  Same as \cref{sec:artifacts:scenarios:2} (\cref{tbl:workflow-config-cost}).
  \\
  \hline%\bottomrule
\end{tabular}
  \caption{Configuration to optimize response time, energy efficiency, and cloud cost.}
  \label{tbl:workflow-config-time-energy-cost}
  }%
\end{table}

%===========================================================
%===========================================================

\subsection{Artifacts and Tools}
\label{sec:artifacts:details}

This section describes CHESS's artifacts and tools, all of which are easily obtained \cite{chess-tools-url}.

%=======================================
%=======================================

\subsubsection{Co-design tools for hybrid workflows}

\paragraph{DataLife}
Paper: \cite{Lee+:2023-SC:data-flow-lifecycles}, URL: \cite{datalife-www}.
%
% DataLife: DataLife is a measurement and analysis toolset for distributed scientific workflows that use I/O and storage for task composition.
%
Distributed scientific workflows pass information – often large volumes – along chains of different computational tasks [in the experiment-instrument-theory cycle], causing data flow bottlenecks in storage and networks. We have developed DataLife, a measurement and analysis toolset for these workflows. DataLife performs data flow lifecycle (DFL) analysis to guide decisions regarding coordinating task and data flows on distributed resources. DataLife provides tools for measuring, analyzing, visualizing, and estimating the severity of flow bottlenecks.

DataLife's measurement introduces techniques that deliver high precision while also imposing minimal overhead. The bottleneck estimator provides several analyses and visualizations to identify and rank opportunities for improving task and data placement and resource assignment. Our technical paper in SuperComputing 2023 introduces data flow lifecycle analysis and describes the fundamental concepts of our methodology and results. With our methodology, we demonstrated the ability to improve the response time of three interesting workflows by 15×, 1.9×, and 10-30×. Leveraging this foundation, we are actively developing other research ideas.

\paragraph{DaYu}
Paper: \cite{Tang+:2024:CLUSTER-dayu}, URL: \cite{dayu-www}.
The combination of ever-growing scientific datasets and distributed workflow complexity creates I/O performance bottlenecks due to data volume, velocity, and variety. Although the increasing use of descriptive data formats (e.g., HDF5, netCDF) helps organize these datasets, it also creates obscure bottlenecks due to the need to translate high level operations into file addresses and then into low-level I/O operations.

DaYu is a method and toolset for analyzing (a) semantic relationships between logical datasets and file addresses, (b) how dataset operations translate into I/O, and (c) the combination across entire workflows. DaYu's analysis and visualization enables identification of critical bottlenecks and reasoning about remediation. We describe our methodology and propose optimization guidelines. Evaluation on scientific workflows demonstrates up to 3.7x performance improvements in I/O time for obscure bottlenecks. The time and storage overhead for DaYu's time-ordered data is typically under 0.2\% of runtime and 0.25\% of data volume, respectively.

\paragraph{FastFlow}
Paper: \cite{Guo+:20xx:fast-flow}.
When distributed scientific workflows are not intelligently executed, they can fail time constraints. To improve workflow response time, FastFlow is a new method of scheduling and resource assignment based on a monitor-analyze-optimize strategy. The key insight is to use the global perspective of interacting critical flows to guide a fast (locally greedy) scheduler that uses data flow projections to select between the better of flow parallelism and flow locality. Our method is linear time, unlike the next-best methods.

\paragraph{Scientific workflows}
A suite of distributed scientific workflows with a range of workload characteristics

%=======================================
%=======================================

\subsubsection{Co-designed workflow modules}

% Interpretation of scientific data streams

\paragraph{SAM-I-Am}
Paper: \cite{Abebe+:2025:SamIAm-semantic-boosting,Abebe+:20xx:SamIAm-semantic-boosting}, URL: \cite{samiam-www}, Demo: \cite{samiam-demo}.
Image segmentation is a critical enabler for tasks ranging from medical diagnostics to autonomous driving.
However, the correct segmentation semantics --- where are boundaries located? what segments are logically similar? --- change depending on the domain, such that state-of-the-art foundation models can generate meaningless and incorrect results. 
Moreover, in certain domains, fine-tuning and retraining techniques are infeasible:
  obtaining labels is costly and time-consuming;
  domain images (micrographs) can be exponentially diverse; and
  data sharing  (for third-party retraining) is restricted.
  To enable rapid adaptation of the best segmentation technology, we propose the concept of \emph{semantic boosting}: given a zero-shot foundation model, \emph{guide} its segmentation and adjust results to match domain expectations.
We apply semantic boosting to the Segment Anything Model (SAM) to obtain \emph{microstructure segmentation} for transmission electron microscopy.
Our booster, SAM-I-Am, extracts geometric and textural features of various intermediate masks to perform mask removal and mask merging operations.
We demonstrate a zero-shot performance increase of (absolute) +21.35\%, +12.6\%, +5.27\% in mean IoU, and a -9.91\%, -18.42\%, -4.06\% drop in mean false positive masks across images of three difficulty classes over vanilla SAM (ViT-L).

\paragraph{SAM-I-Am-LabelStudio}
URL: \cite{samiam-labelstudio-www}.
It is critical to pre-train AI models with accurate labels.
However, precisely segmenting and labeling an image dataset is extraordinarily time-consuming.
We created a SAM-I-Am-assisted labeling assistant within LabelStudio to to rapidly create precise labels for image segments of identical materials.
Before using this tool we could only obtain 3 labeled images in several hours; with it, we obtained 50 labeled images in under 2 hours.

\paragraph{SAM-I-Am Dataset}
Dataset: \cite{samiam-dataset}.
A dataset consisting of labels for 82 scanning TEM images, (mostly 1024 $\times$ 1024) but some with (2048 $\times$ 2048) pixel resolution, from two domain experts.
Each image contains 2--4 materials from among compounds including La:SrTiO$_3$, SrTiO$_3$, Ge, Nb:SrTiO$_3$, WO$_3$, LaFeO$_3$, LaMnO$_3$, La$_{0.8}$Sr$_{0.2}$FeO$_3$, and Pt / C captured in annular dark-field (ADF) and high-angle annular dark-field (HAADF) modes.
Moreover, some images show vacuum regions while some material surfaces show certain defects.
The experts divided the images into 3 difficulty classes based on the intricacy of the texture patterns.

Each image contains 2--4 materials from among compounds including La:SrTiO3, SrTiO3, Ge, Nb:SrTiO3, WO3, LaFeO3, LaMnO3, La0.8Sr0.2FeO3, and Pt / C captured in annular dark-field (ADF) and high-angle annular dark-field (HAADF) modes. Moreover, some images show vacuum regions while some material surfaces show certain defects. As shown in Fig. 3, the experts divided the images into 3 difficulty classes based on the intricacy of the texture patterns.

%Distributed model training

\paragraph{MassiveGNN}
Paper: \cite{Sarkar+:2024:CLUSTER-massive-gnn}, URL: \cite{massivegnn-www}.
%
% Performant and productive training for massively connected (distributed) GNNs within Deep Graph Library.
%
Graph Neural Networks (GNN) are indispensable in learning from graph-structured data, yet their rising computational costs, especially on massively connected graphs, pose significant challenges in terms of execution performance. However, approaches requiring a distributed-memory graph usually suffer from communication overhead and load imbalance due to non-determinism in standard training methods.

MassiveGNN introduces performant and productive training for massively connected (distributed) GNNs within the state-of-the-art Amazon DistDGL distributed GNN framework. It brings practical trade-offs for improving the sampling and communication overheads for representation learning on distributed graphs by developing a parameterized prefetch and eviction scheme. It uses a continuous prefetch and eviction scheme to tally and move nodes (features) into and out of a preallocated buffer For the next minibatch, all while concurrently learning on the current minibatch.
On GraphSAGE architecture we demonstrate 15--40\% improvement in end-to-end training performance on the NERSC Perlmutter supercomputer for various OGB datasets.

%Compression for science:

\paragraph{ViSemZ}
Paper: \cite{Zhang+:20xx:visemz-semantic-compression}.
Scientific images are crucial for many experimental sciences, but dataset volumes pose significant challenges. Effective image compression must be quick, achieve high ratios, and enable automated interpretation by preserving essential domain features. Traditional image compressors like JPEG can distort critical textures at high compression ratios. In contrast, AI-based compression offers excellent image quality and impressive ratios. However, they are much slower than traditional approaches. To address this, ViSemZ is a high-performance image compressor that preserves \emph{vi}sual \emph{sem}antics.
Evaluations on general and scientific datasets demonstrate that ViSemZ improves the overall compression throughput by up to 9.2$\times$ without compromising visual semantics, meanwhile maintaining high compression ratio, effectively bridging the gap between traditional and AI-based compression speeds.

\paragraph{PRISMScope}
URL: \cite{prismscope-www}
Given a description of the material's crystalline structure and microscope parameters, PRISMScope generates a range of 2D image projections that are associated with a specific combination of parameters.
The results enables association between a material, its crystalline structure, and STEM settings.
Based on the input request, PRISMScope automatically retrieves a set of relevant parameters from the Materials Project
% \url{https://materialsproject.org}
and then uses Prismatic,
% Prismatic: https://prism-em.com
which employs a multislice method to compute a 4D scattering matrix.
PRISMScope then generates the desired projections to obtain a 2D STEM image.
% Prismatic multislice is a more general version of PRISM multislice

\paragraph{PyChip ensemble trainer}
Ensemble-based training for AutoEM's PyChip material classifier.

%============================================================================
%============================================================================

\section{Workflow frameworks: Nextflow and SkyPilot}
\label{sec:frameworks}

%===========================================================
%===========================================================

\subsection{Nextflow}
\label{sec:nextflow-intro}

Nextflow~\cite{nextflow-url} is a workflow system for creating scalable, portable, and reproducible workflows.
There are a myriad of options for coordinating workflows.
However, if the short- and long-term needs of research and open science are considered, usually only a handful remain for consideration.
The CHESS project desired to select on sample coordination system that satisfied the following criteria:
\begin{myitemize}
\item productivity, or expressiveness of the language. With NextFlow there is a simple and direct translation path for existing scripts and methods on facility resources. At the same time, it is possible to incrementally optimize the task coordination and dataflow using techniques like fine-grained pipelining and lightweight tasks. The results will still be compatible with virtual platforms composed of facility and cloud resources.
  
\item portability, i.e., hybrid execution across multiple cloud providers;
\item performance, enabling precise control and algorithm specializations
\item maturity: robustness, many years of development, and potentially the availability of professional support
\item suitability for science and familiarity of paradigms: design philosphy based supporting a scientific environment
\item openness: a framework that permitts full (100\%) ownership by scientists meaning there is lock by a specific cloud vendor; open source implementations, potentially dispersed across scientific collaborations, 
\end{myitemize}
In our evaluation period, NextFlow emerged as the top contendor.

NextFlow is based on the dataflow programming model, which greatly simplifies the writing of parallel and distributed pipelines, allowing you to focus on the flow of data and computation.
Nextflow can deploy workflows on a variety of execution platforms, including your local machine, HPC schedulers, AWS Batch, Azure Batch, Google Cloud Batch, and Kubernetes.
Additionally, it supports many ways to manage your software dependencies, including Conda, Spack, Docker, Podman, Singularity, and more.

%\url{https://www.nextflow.io/docs/latest/}

%===========================================================
%===========================================================

\subsection{Nextflow and SkyPilot}
\label{sec:skypilot-intro}

Nextflow can manage cloud launches by itself or by deferring to SkyPilot~\cite{skypilot-url,skypilot}.
The motivation for SkyPilot is that the availability of computational resources (especially high-demand GPUs) can differ by cloud provider.
Further pricing, especially for spot, can differ by not only by cloud provider but within the different regions of the same vendor.

SkyPilot is an inter-cloud broker that can select the best resources both among multiple cloud providers and within each cloud provider, optimizing for time and price.
SkyPilot manages a current database of availability and costs across cloud and within each cloud's region.
When defining a job (workflow or task) with SkyPilot, its resources are specified as requirements, i.e., in a declarative (vs. imperative) fashon.
An optimization objectives (time and/or cost) can also be specified.
When a job is launched, SkyPilot consults the current database and  selects the resources that best meet the objective.
SkyPilot can also dynamically transition long-running jobs from one cloud region/provider to another.

%============================================================================
%============================================================================

\clearpage

\pagestyle{empty}% remove page numbers

\renewcommand{\refname}{\mychaptertitle{References Cited}}

\bibliographystyle{\mybibstyle}

\bibliography{refs}

\end{document}